\documentclass[reprint,
superscriptaddress,
nobibnotes,
amsmath,amssymb,
aps,
longbibliography
]{revtex4-2}

\usepackage{dcolumn}
\usepackage{amssymb, esint}
\usepackage{physics}
\usepackage{bm}
\usepackage{graphicx}
\usepackage[colorlinks,allcolors=blue]{hyperref}
\usepackage{url}
\begin{document}

%\title{Classification of periodically-driven Hamiltonians in the adiabatic limit}

\title{A time-based Chern number in periodically-driven systems in the adiabatic limit}

\author{I-Te Lu}
\affiliation{Max Planck Institute for the Structure and Dynamics of Matter and Center for Free-Electron Laser Science, Luruper Chaussee 149, 22761, Hamburg, Germany}

\author{Dongbin Shin}
\affiliation{Max Planck Institute for the Structure and Dynamics of Matter and Center for Free-Electron Laser Science, Luruper Chaussee 149, 22761, Hamburg, Germany}

\author{Umberto De Giovannini}
\affiliation{Max Planck Institute for the Structure and Dynamics of Matter and Center for Free-Electron Laser Science, Luruper Chaussee 149, 22761, Hamburg, Germany}
\affiliation{Università degli Studi di Palermo, Dipartimento di Fisica e Chimica—Emilio Segrè, Palermo I-90123, Italy}

\author{Hannes Hübener}
\affiliation{Max Planck Institute for the Structure and Dynamics of Matter and Center for Free-Electron Laser Science, Luruper Chaussee 149, 22761, Hamburg, Germany}

\author{Jin Zhang}
\affiliation{Max Planck Institute for the Structure and Dynamics of Matter and Center for Free-Electron Laser Science, Luruper Chaussee 149, 22761, Hamburg, Germany}

\author{Simone Latini}
\affiliation{Max Planck Institute for the Structure and Dynamics of Matter and Center for Free-Electron Laser Science, Luruper Chaussee 149, 22761, Hamburg, Germany}

\author{Angel Rubio}
\email{angel.rubio@mpsd.mpg.de}
\affiliation{Max Planck Institute for the Structure and Dynamics of Matter and Center for Free-Electron Laser Science, Luruper Chaussee 149, 22761, Hamburg, Germany}
\affiliation{Center for Computational Quantum Physics (CCQ), The Flatiron Institute, 162 Fifth avenue, New York NY 10010, United States}

%\date{\today}

%%%%%%%%%
% Abstract
%%%%%%%%%
\begin{abstract}
%
%To define the topology of driven systems, recent works have proposed synthetic dimensions to uncover the underlying parameter space of topological invariants like the $2$D Chern number. 
To define the topology of driven systems, recent works have proposed synthetic dimensions as a way to uncover the underlying parameter space of topological invariants.
%
% Such synthetic dimensions are often defined from time-related quantities (e.g., the energy lattice structure formed by the harmonics of a driving frequency) but not by time itself.
% %
%
Using time as a synthetic dimension, together with a momentum dimension, gives access to a \textit{synthetic} $2$D Chern number.
It is, however, still unclear how the synthetic 2D Chern number is related to the Chern number that is defined from a parametric variable that evolves with time.
%
%It is, however, still unclear how directly to use time itself as a synthetic dimension.
%
Here we show that in periodically driven systems in the \textit{adiabatic limit}, the synthetic $2$D Chern number is a multiple of the Chern number defined from the parametric variable. 
The synthetic 2D Chern number can thus be engineered via how the parametric variable evolves in its own space.
%This integer multiple can thus be used to classify the evolution of a periodically-driven system in the adiabatic limit.
%, i.e., without gap closing. 
%
We justify our claims by investigating Thouless pumping in two $1$D tight-binding models, a three-site chain model and a two-$1$D-sliding-chains model. 
%
%We show that to use time as a synthetic dimension, one needs to take into account the underlying parametric variable that is coupled to time in the Hamiltonian.
%
The present findings could be extended to higher dimensions and other periodically driven configurations.
\end{abstract}

\maketitle

% Introduction
%\section{Introduction}\label{sec:intro_sec}

% broad introduction 
Periodically-driven quantum systems under an external time-dependent driving source have emerged as a platform for generating new exotic matter states in a wide range of materials such as atomic gases, $2$D, twisted, and bulk materials~\cite{oka.kitamura_2019,giovannini.hubener_2019,topp.jotzu.ea_2019a,rudner.lindner_2020,weitenberg.simonet_2021}. 
%
%These periodically-driven 
These systems can be interpreted in terms of quasiparticles that are modified or dressed by the harmonics of external driving sources from low to high frequencies~\cite{bukov.dalessio.ea_2015a,giovannini.hubener_2019,rodriguez-vega.vogl.ea_2021a}.
%
%The quasiparticles band structures can in turn be quantified or classified by the band topological invariants~\cite{rudner.lindner_2020}. 
The quasiparticles band structures can in turn be classified by the band topological invariants~\cite{rudner.lindner_2020}.

The 2D Chern number, a topological invariant on a closed surface, is usually evaluated in the space-related dimensions, e.g. Brillouin zone, of solid state materials, which are limited at most to the $3$D spatial space. 
To extend the classification to dimensions other than spatial dimensions, the so-called synthetic (non-spatial) dimensions have been proposed to expand the concepts of topology to higher dimensions, opening a new way to design materials properties~\cite{jian.xu_2018,lohse.schweizer.ea_2018a,ozawa.price_2019,boyers.crowley.ea_2020,lustig.lustig.ea_2021,chen.zhang.ea_2021}.

For periodically-driven systems, one of the relevant synthetic dimensions is the Floquet frequencies of an external driving; based on the topological invariants of the synthetic dimensions, a number of interesting physical phenomena such as quantized energy exchange between the external driving fields have been predicted~\cite{martin.refael.ea_2017, baum.refael_2018,peng.refael_2018,crowley.martin.ea_2019,qi.refael.ea_2021}. 
In contrast to such a dimension formed by the Floquet frequencies, using time itself directly as a synthetic dimension is somewhat less intuitive, and this is what we are going to explore in the present work.
%

%Nevertheless, time has been used as a synthetic dimension, together with the momentum dimension, to compute a $2$D Chern number on the two dimensional plane for a time- and spatially-periodic Hamiltonian $H(t+T, x+a)=H(t, x)$, where $t$ is time, $T$ the time period, $x$ the spatial coordinate, and $a$ the lattice constant~\cite{nakajima.tomita.ea_2016,lohse.schweizer.ea_2016}. 
%
Indeed, causality sets that we cannot move backward in time to define a closed loop in parameter space that is needed when computing a $2$D Chern number that involves the time dimension. 
Nevertheless, time has been used as a synthetic dimension, together with the momentum dimension, to compute a $2$D Chern number for a time- and spatially-periodic Hamiltonian $H(t+T, x+a)=H(t, x)$, where $t$ is time, $T$ the time period, $x$ the spatial coordinate, and $a$ the lattice constant~\cite{nakajima.tomita.ea_2016}.
In such cases, the Chern number represents the number of pumped quantum charges, known as Thouless (or quantum) pumping~\cite{thouless_1983}, which has been demonstrated in many different experiments~\cite{nakajima.tomita.ea_2016,lohse.schweizer.ea_2016,ma.zhou.ea_2018}.
%in real space following the adiabatic perturbation
%
% gap sentence

%However, the Hamiltonian of a periodically-driven system is not directly coupled to time; instead, it is coupled to a physical parameter that enters the Hamiltonian at a given time $t$, for example, $H[\alpha(t),x]$, where $\alpha(t)$ is a parametric function that evolves with time, e.g. the position of the nuclei in a solid, the vector potential of an external field, etc.
%

In this work we reveal that in a periodically driven system, even though the time dependence is parametric, the \textit{synthetic} 2D Chern number$-$obtained using time as one of the two dimensions$-$can be built upon another Chern number defined on the parametric dimension $\alpha$, which evolves with time and can define a loop in its parameter space, e.g., the position of the nuclei in a solid, the vector potential of an external field, etc.
%
%Here we demonstrate that if one evaluates a synthetic $2$D Chern number that involves time as one of the two dimensions, the synthetic Chern number should reflect another Chern number defined on the parametric dimension. 
% action 
We provide an analytical proof that, in the \textit{adiabatic limit} the $2$D Chern number, $C^{\{t-k\}}$, which involves the time $t$-dimension and a spatial (momentum) $k$-dimension, is always an integer multiple of another, more fundamental 2D Chern number, $C^{\{\alpha-k\}}$, which involves a periodic parametric function $\alpha(t)$ that follows $\alpha(t+T) = \alpha(t)+l$ where $l$ is an integer. 
%
% resolution
This integer multiple can thus be used to classify a periodically-driven system in the adiabatic limit. 
Our finding suggests that the proposed synthetic 2D Chern number, $C^{\{t-k\}}$, can be designed via the trajectory of the parametric variable $\alpha(t)$.
%
%Below we use two tight-binding models, an infinite three-sites chain and a two-$1$D-sliding-chains model free to move with respect to each other, to demonstrate the proposed formula. 
%
%This work shows that topological quantities that involve the time dimension are determined by the underlying physical parameters that directly couples to time. 
%

\section*{Time as a synthetic dimension}\label{sec:method_sec}

%%%%%%%%%%%%%%%%%%%%%%%
% theorem introduction 
%%%%%%%%%%%%%%%%%%%%%%%
For the sake of simplicity and without loss of generality, suppose we have a $1$D real-space periodic Hamiltonian $H(\alpha,x+a)=H(\alpha,x)$ with a tuning time-dependent parameter $\alpha(t)$; here $a$ is set to $2\pi$ below without loss of generality.
%
%Due to the periodicity in real space, 
It is convenient to transform the Hamiltonian to the momentum space, $H(\alpha,k)$, with the crystal momentum $k$ in the first Brillouin zone. 
Starting from such Hamiltonian, we make the following assumptions. 
%
%We assume, without loss of generality, that the real-space periodicity determined by the lattice constant $a$ is $2\pi$.
%; that is, $H(\alpha,x+2\pi)=H(\alpha,x)$. 
% 
The Hamiltonian is periodic in the $\alpha$ parametric space, $H(\alpha+1,k)=H(\alpha,k)$; we are allowed to do so as $\alpha$ can always be conveniently rescaled. 
We further assume that no gap closure occurs$-$adiabatic approximation$-$for any value of $\alpha$, enabling us to compute the Chern number later; we will address the  interesting  case of gap closing in a follow up work.
For a given $\alpha$ value, the eigenstates for the Hamiltonian $H(\alpha,k)$ are the Bloch waves with a band index $n$, $\ket{u_{n}(\alpha,k)}$. 
Below we use the periodic gauge for the Bloch wave functions, i.e., $\ket{u_{n}(\alpha,k+1)}=\ket{u_{n}(\alpha,k)}$~\cite{vanderbilt2018berry}.
The Chern number for the Bloch wave function with band index $n$ on the $\alpha-k$ plane is computed using the standard definition~\cite{vanderbilt2018berry}: $C_{n}^{\{\alpha-k\}}=
\frac{1}{2\pi}\oiint_{S_{\alpha k}}\Omega_{n,\alpha k}\ d\alpha dk$,
% %
% \begin{equation}
% C_{n}^{\{\alpha-k\}}=
% \frac{1}{2\pi}\oiint_{S_{\alpha k}}\Omega_{n,\alpha k}\ d\alpha dk,
% \end{equation}
% %
where the integration is over the $2$D closed surface $S_{\alpha k}$ formed by the two dimensions, $\alpha$ and $k$. 
The Berry curvature is then defined as $\Omega_{n,\alpha k} = \partial_{\alpha}A_{n, k}-\partial_{k}A_{n, \alpha}$, 
% %
% \begin{equation}
% \Omega_{n,\alpha k} = \partial_{\alpha}A_{n, k}-\partial_{k}A_{n, \alpha},
% \end{equation}
% %
where $\partial_{\alpha}=\partial/\partial \alpha$, $\partial_{k}=\partial/\partial k$, and the Berry connection $A_{n, \alpha/k} = \mel{u_{n}(\alpha,k)}{\partial_{\alpha/k}}{u_{n}(\alpha,k)}$.

%%%%%%%%%%%%%%%%%%%%%%%%%%%%%%%
% Introduce time into the alpha value   
%%%%%%%%%%%%%%%%%%%%%%%%%%%%%%%
Now we explicitly consider the time periodicity in the Hamiltonian $H(t+T)=H(t)$ under the assumption that the time-dependent change is adiabatic at each time.
To make sure that the Hamiltonian is invariant after one $T$-period, the path of the parametric variable $\alpha(t)$ must satisfy $\alpha(t+T)=\alpha(t)+l$, where $l$ is an integer, such that $H[\alpha(t+T)]=H[\alpha(t)+l]=H[\alpha(t)]$.
Here we state our main result: \textit{The Chern number on the $t-k$ plane, $C_{n}^{\{t-k\}}$, is a multiple of that evaluated on the $\alpha-k$ plane, i.e., $C_{n}^{\{t-k\}} = lC_{n}^{\{\alpha-k\}}$}.
The implications of the connection between the two invariants are as follows: i) a periodically-driven Hamiltonian in the adiabatic limit can be categorized into different groups based on the integer $l$;
%
%2) a periodically-driven system with different paths $\alpha(t)$ belongs to the same category, if those paths have the same integer $l$ to satisfy $\alpha(t+T)=\alpha(t)+i$;
%
ii) the Chern number on the $\alpha-k$ plane, $C_{n}^{\{\alpha-k\}}$, forms the building block for the Chern number on the $t-k$ plane, $C_{n}^{\{t-k\}}$~\footnote{The Chern number on the $t-k$ plane can be determined from that on the $\alpha-k$ plane in the adiabatic limit; therefore, we say the latter is more fundamental.}.
%, where $\alpha$ is the parameter that directly couples to time. 
%
% Because of the integer shown in the theorem, the Chern number on the $\alpha-k$ plane is \textit{enhanced} due to the periodical driving. 
% %
% Thus, we call the above theorem as the time-periodically enhanced Chern number theorem.
The physical meaning of the integer $l$ is how many copies of the Chern number $C_{n}^{\{\alpha-k\}}$ the trajectory $\alpha(t)$ picks up within a time period.

%%%%%%%%%%%%%%%%%%%%%%%%%%%%%%%
% Proof of the theorem  
%%%%%%%%%%%%%%%%%%%%%%%%%%%%%%%
% start with the t-k Chern number, and use the Berry connection to connect to the alpha-k Chern number
Here we prove the above statement as follows:
\begin{equation}\label{eq:ctk2cak}
\begin{aligned}
C^{\{t-k\}}_{n}& =\frac{1}{2\pi}\int_{0}^{T}dt\int_{0}^{1}dk\ \Omega_{n,tk} =\frac{1}{2\pi}\int_{0}^{T}dt\int_{0}^{1}dk\ \partial_{t} A_{n, k} \\ & = \frac{1}{2\pi} \int_{0}^{T}dt\  \partial_{t}\left(\int_{0}^{1}dk\ A_{n, k}\right) =\frac{1}{2\pi} \int_{0}^{T}dt\  \partial_{t}\phi_{n}^{(k)}  \\ & = \frac{1}{2\pi} \int_{\alpha(0)}^{\alpha(T)}d\alpha\ \partial_{\alpha}\phi_{n}^{(k)} \\ & = l\times \left[\frac{1}{2\pi}\int_{\alpha(0)}^{\alpha(0)+1}d\alpha\ \partial_{\alpha}\phi_{n}^{(k)}\right] = l C^{\{\alpha-k\}}_{n}.
\end{aligned}
\end{equation}
For the last equality in the first line, the integration of $\partial_{k}A_{t}$ over the $k$ coordinate vanishes due to the periodic gauge for the Bloch wave function. 
At the end of the second line, we define the Berry phase along the $k$ coordinate as $\phi_{n}^{(k)}$.
In the third line, we change the variable from $t$ to $\alpha$. 
In the last line, the integration interval $[\alpha(0), \alpha(0)+l]$ can be exactly separated into $l$ intervals, i.e., $[\alpha(0), \alpha(0)+1]$,..., $[\alpha(0)+l-1, \alpha(0)+l]$, for each of which the Chern number $C_{n}^{\{\alpha-k\}}$ is the same. 
In $1$D, the Berry phase $\phi_{n}^{(k)}$ is related to the center of the Wannier function via the formula $a\phi_{n}^{(k)}/2\pi$~\cite{vanderbilt2018berry}.%, where $a$ is the lattice constant.
The center can be regarded as the position of the charge, which may move after \textit{one parametric loop}, i.e., quantum pumping.
The number of pumped charges is equal to the Chern number $C_{n}^{\{\alpha-k\}}$ obtained on the $\alpha-k$ plane.

%%%%%%%%%%%%%%%%%%%%%%%
%%%%%%  Figure 1 %%%%%%
%%%%%%%%%%%%%%%%%%%%%%%
%
\begin{figure}[b]
\centering
\includegraphics[width=\linewidth]{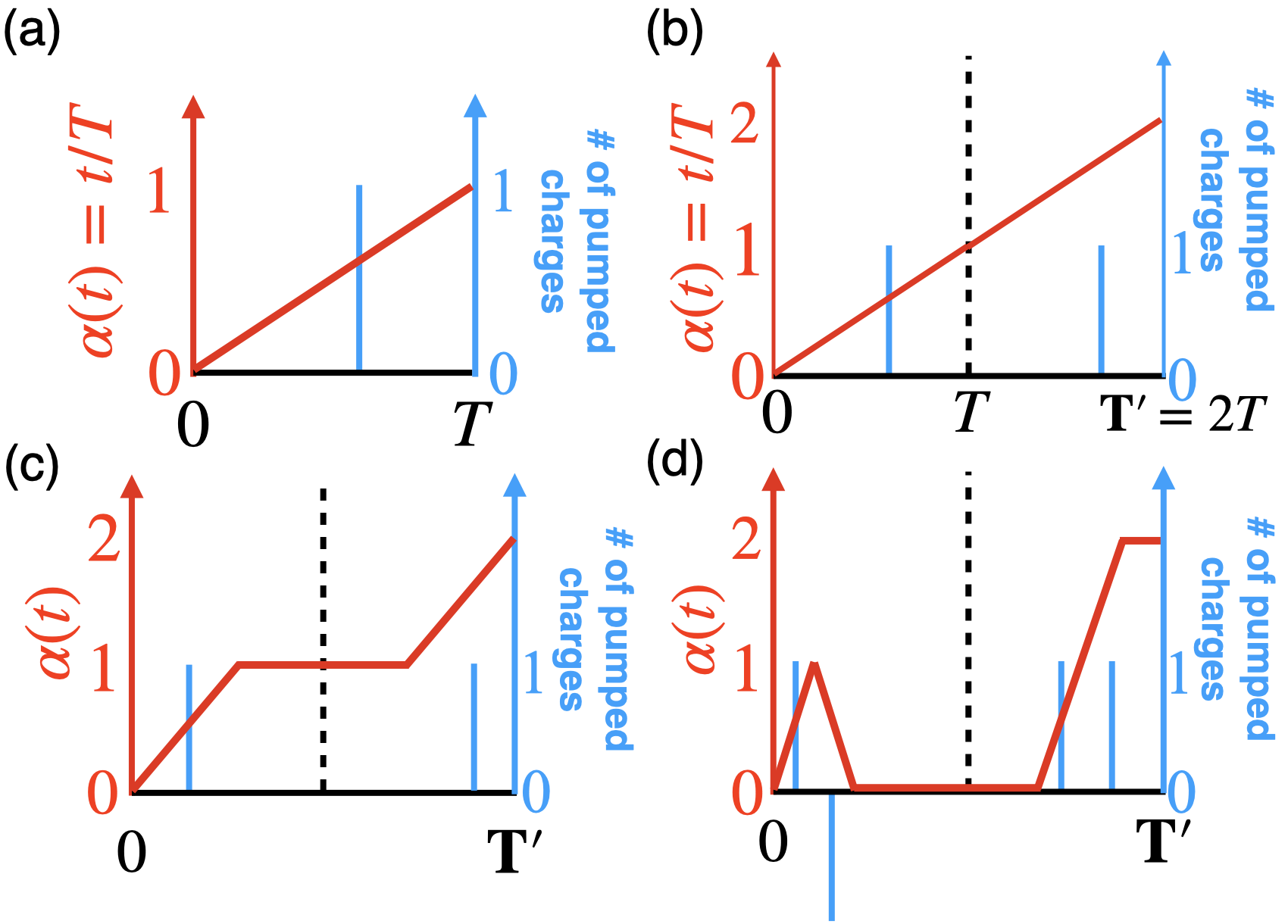}
\caption{
Schematic illustration of the formula $C^{\{t-k\}}_{n}= l C^{\{\alpha-k\}}_{n}$.
(a) The path for the parametric variable $\alpha(t)=t/T$ where $T$ is the period. 
(b) The extension of (a) with a newly defined period $T'=2T$.
(c) and (d) Two deformed parametric paths and the corresponding number of pumped charges as a function of time within the new time period $T'$.
}
\label{fig:rescaleT}
\end{figure}

With the general concept of quantum pumping in mind, Fig.~\ref{fig:rescaleT} provides a conceptual illustration of the implications of Eq.~(\ref{eq:ctk2cak}) and the meaning of the integer $l$.
We chose $\alpha(t)=t/T$ such that the time dimension is equivalent to the parametric dimension. 
Fig.~\ref{fig:rescaleT}(a) shows the parametric variable and the number of pumped charges as a function of time, assuming a Chern number $C^{\{\alpha-k\}}=C^{\{T-k\}}=1$; 
that is, the number of pumped charges is $+1$ when $\alpha$ goes from $0$ to $1$, while $-1$ along the opposite direction.
We now consider the same periodically driven system but extend its period to $2T$. 
We use the same parametric path and redefine the period as $T'=2T$, as shown in Fig.~\ref{fig:rescaleT}(b). 
In this case the number of pumped charges, or $C^{\{T'-k\}}$, becomes $2$. 
Deforming the parametric path in this new period while fixing the initial and final values of $\alpha$ like the ones shown in Fig.~\ref{fig:rescaleT}(c) and (d) does not affect the Chern number $C^{\{T'-k\}}$, which  remains $2$. 
This example illustrates the topological nature of the formulation and how  the integer $l$ naturally emerges from the extension of the old period and the redefinition of a new period. 
In addition, if we use the number of pumped charges during one period as a classification approach, any periodically-driven system with the same number of pumped charges within the adiabatic limit belong to the same group.
On the other hand, if we classify the system in terms of the topology hidden in the parametric function $\alpha$, systems with different periodicity in the driving can still be classified as in the same class in the $\alpha-k$ plane.

%\section{Results and Discussion}\label{sec:res_dis_sec}

Here we use two $1$D-spatially-periodic lattice models to demonstrate our above claims.
%unveil the underlying topological classification in the adiabatic time limit. 
%
The first example is an infinite $1$D periodic tight-binding chain with three sites in a unit cell, while the second one consists of two-parallel-$1$D tight-binding chains described by two atoms in a unit cell.
Below we use the code PythTb~\cite{pythtb-web} to construct the tight-binding Hamiltonians and compute the Berry phase, Berry curvature, and Chern number.
The details on the parameters for the Hamiltonians can be found below.
%
%Note that we do not run any dynamics calculations. 

%%%%%%%%%%%%%%%%%%%%%%%
%%%%%%  Figure 2 %%%%%%
%%%%%%%%%%%%%%%%%%%%%%%
%
\begin{figure}[b]
\centering
\includegraphics[width=\linewidth]{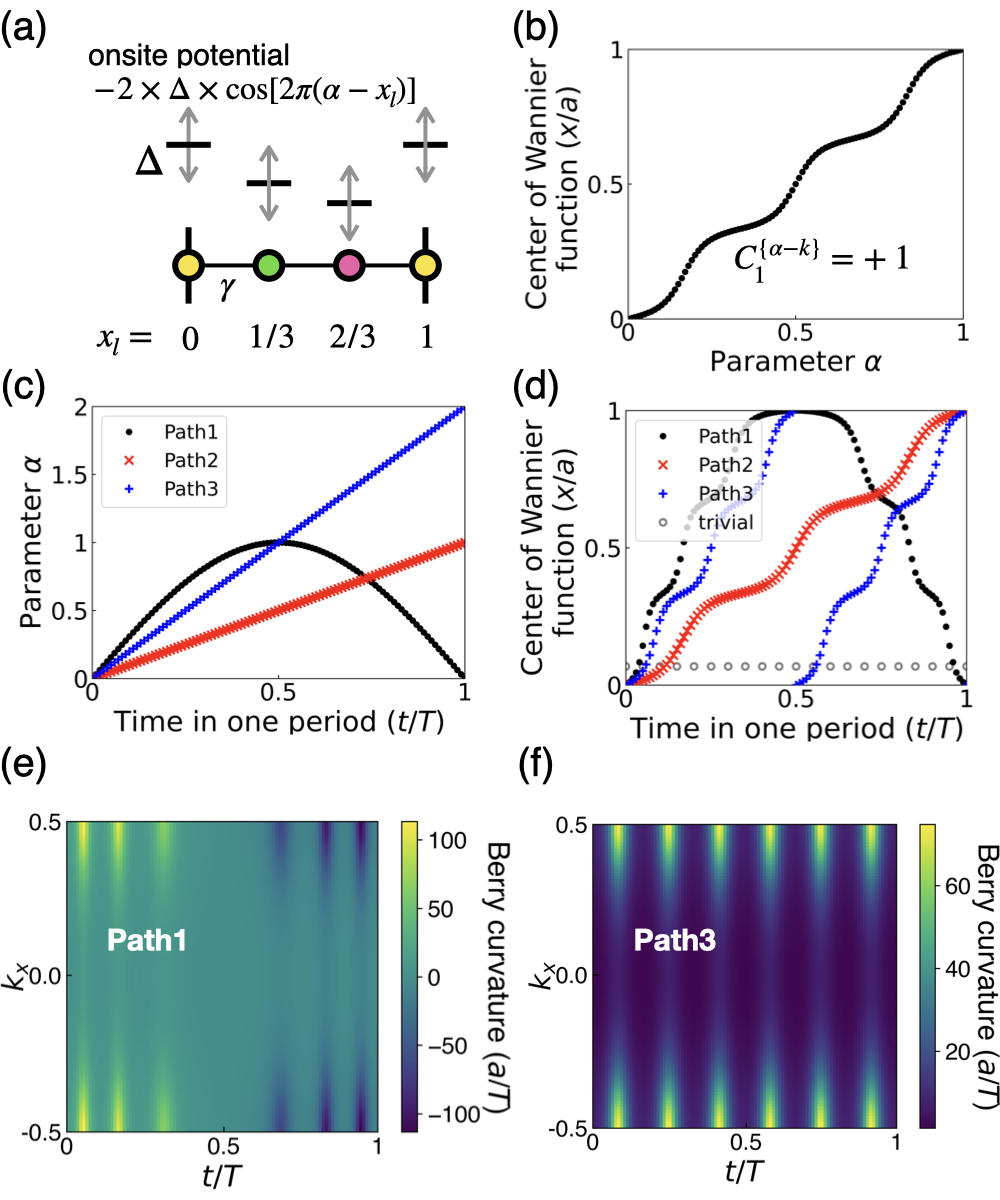}
\caption{
Bulk 1D periodic three-site tight-binding model with a non-trivial quantum pumping.
(a) The unit cell with the parameters described in the main text. 
(b) The evolution of the center of the Wannier function for the lowest band (labelled by $1$) along a loop of the parametric variable $\alpha$.  
The Chern number on the $\alpha-k$ plane, $C_{1}^{\{\alpha-k\}}$, is $1$.
(c) Three chosen paths for $\alpha(t)$. 
Path1: $\alpha(t)=\sin^{2}(\pi t/T)$; Path2: $\alpha(t)=t/T$; Path3: $\alpha(t)=2t/T$.
(d) The evolution of the center of the Wannier function for each path within one period. 
For comparison, we provide 
a trivial case [see Eq.~(\ref{eq:trivial_H})], of which the Wannier function remains a constant, in this case, $0.07$.
(e) and (f) The Berry curvature on the $t-k$ plane for Path1 ($C_{1}^{\{t-k\}}=0$) and Path3 ($C_{1}^{\{t-k\}}=2$), respectively.
%
%The color map shows the magnitude of the Berry curvature; the brighter the color is, the larger the Berry curvature is.
%
%Here we use $\Delta=2.0$ and $\gamma=-1.0$ for the model.
}
\label{fig:3site_nontrivial}
\end{figure}

\subsection*{Three-site 1D tight-binding model}
We use a standard $1$D three-site model shown in Fig.~\ref{fig:3site_nontrivial}(a), a simple toy model used for explaining the fundamental concepts of topology, e.g., Berry phase and Chern number, in Ref.~\cite{vanderbilt2018berry}.
The Hamiltonian is
\begin{equation}\label{eq:3site_model}
%\begin{aligned}
    H(\alpha)=-2\Delta\sum_{i}\cos[2\pi(\alpha-x_{i})]c_{i}^{\dagger}c_{i}+\gamma\sum_{i}\left[c_{i+1}^{\dagger}c_{i}+\rm{c.c.}\right],
%\end{aligned}
\end{equation}
%
%where the first term describes the onsite potential and the second the coupling between the nearest neighbors.
%
where $\Delta$ is the onsite-potential amplitude, $\alpha$ the parametric variable to control the phase of each site, $x_{i}$ the atomic position in terms of the lattice constant $a$, $\gamma$ the coupling constant between the nearest neighbors, and $c_{i}$($c_{i}^{\dagger}$) the annihilation (creation) operator for an electron at site $i$. 
To make the system insulating for any $\alpha$ value, we use $\Delta=2.0$ and $\gamma=-1.0$.
As a result, the Hamiltonian has a non-trivial Chern number $C_{n}^{\{\alpha-k\}}$ for each band $n$.
The Chern numbers for all the three bands ($n=1$, $2$ and $3$) on the $\alpha-k$ plane, $C^{\{\alpha-k\}}_{n}$, are $1$, $-2$, and $1$, respectively.
In practice, this model can be realized, e.g., using a series of connected quantum dots, each of which is coupled to an external voltage to control the onsite potential. 
Here we focus on the lowest band$-$i.e., by considering it completely filled$-$but all the features and phenomena discussed below are also observed in the other bands.
Fig.~\ref{fig:3site_nontrivial}(b) shows the center of the Wannier function of the lowest band as a function of $\alpha$; indeed, the number of pumped charge is equal to 1, i.e., $C^{\{\alpha-k\}}_{1}=1$.
%
% The center of the Wannier function in $1D$ is related to the Berry phase $\phi^{(k)}$ via the formula $a\times\phi^{(k)}/2\pi$~\cite{vanderbilt2018berry}, where $a$ is the lattice constant.
% %
% The center can be regarded as the position of the charge, which moves by one lattice constant after \textit{one parametric loop}, i.e., quantum pumping.
% %
% The number of the pumped charge is equal to the Chern number $C_{n=1}^{\{\alpha-k\}}$ obtained on the $\alpha-k$ plane; in this case, it is equal to $+1$.

We now consider the time profile of the parametric variable $\alpha(t)$.
Fig.~\ref{fig:3site_nontrivial}(c) shows three arbitrary paths for $\alpha(t)$ within one $T$-period.
The paths are chosen such that the $\alpha$ value differs by one integer $l$ after one period; this guarantees time-periodicity in the Hamiltonian.
%, because of $H(\alpha+l)=H(\alpha)$.
% 
For Path$1$, $2$, and $3$, the $l$ values are $0$, $1$, $2$, respectively.
The time evolution of the Wannier center for each path in one $T$-period is shown in Fig.~\ref{fig:3site_nontrivial}(d).
If we count the number of net pumped charges at $x = 0.5a$, the number for Path1 (2, 3) is 0 (1, 2). 
The number is equal to the Chern number evaluated on the $t-k$ plane, $C_{n}^{\{t-k\}}$.

Fig.~\ref{fig:3site_nontrivial}(e) and (f) show the Berry curvature on the $t-k$ plane for Path1 and 3, respectively, within one $T$-period. 
The Berry curvature for Path1 shows a symmetric feature: positive and negative values before or after $t/T=0.5$, respectively.
This is because $\alpha(t)$ returns to its original value after one period [see Fig.~\ref{fig:3site_nontrivial}(c)] so does the corresponding center of the Wannier function [see Fig.~\ref{fig:3site_nontrivial}(d)].
Within one $T$-period, both Path1 and Path3 access all the possible Hamiltonians $H(\alpha)$ because of $H(\alpha+1)=H(\alpha)$; in this case, $\alpha$ covers all the value in the interval $[0, 1]$. 
Interestingly, the Berry curvature and the Chern number are different for these two paths: one has a Chern number of $0$, and the other $2$.
For Path2, the value of its Berry curvature is just half of that for Path3 with the time range $[0, 0.5]$ and its Chern number is $1$; this Berry curvature is equivalent to the Berry curvature on the $\alpha-k$ plane because of $\alpha(t)$=t/T.
The results demonstrate that one can create a different topological invariant in the $t-k$ space by designing how the parametric variable evolves in its own space. 
In addition, the Chern number on the $t-k$ plane is exactly equal to the multiplication of the integer attached to each path and the Chern number obtained on the $\alpha-k$ plane. 
This suggests that we can use the integer $l$ to classify a periodically-driven Hamiltonian.

%%%%%%%%%%%%%%%%%%%%%%%
%%%%%%  Figure 3 %%%%%%
%%%%%%%%%%%%%%%%%%%%%%%
%
\begin{figure}[t]
\centering
\includegraphics[width=\linewidth]{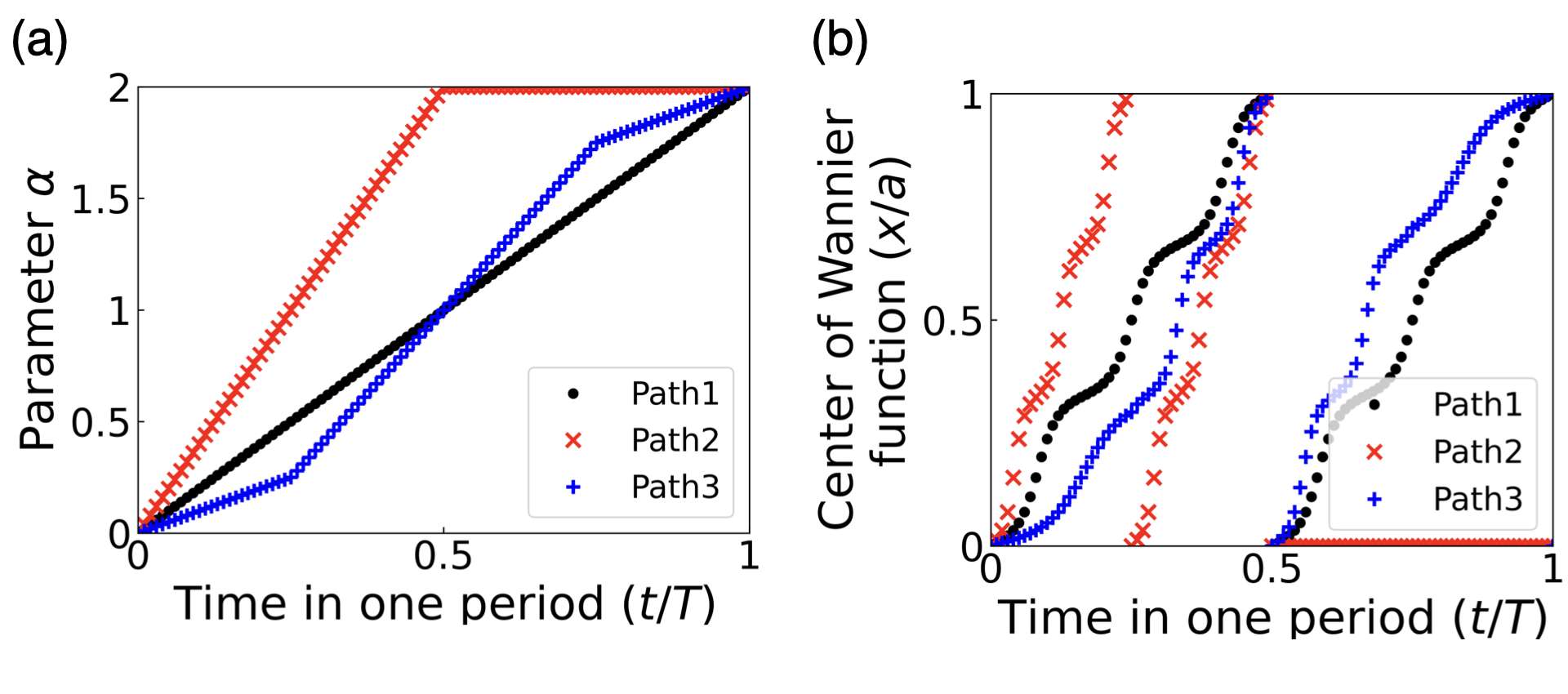}
\caption{
Deformation of the parametric paths with the same Chern number on the $t-k$ space.
(a) Three paths with the same initial value $\alpha(t=0)=0$ and the same final value $\alpha(t=T)=2$ within one period. 
(b) The evolution of the center of the Wannier function for each path within one period. 
}
\label{fig:3site_deformpath}
\end{figure}

Based on Eq.~(\ref{eq:ctk2cak}), the Chern number evaluated on the $t-k$ plane, $C_{1}^{\{t-k\}}$, does not change, if we deform the parametric paths by keeping the same starting and ending value within one $T$-period.
Fig.~\ref{fig:3site_deformpath}(a) shows three possible such paths. %with the same starting and ending $\alpha$ value within one period.  
The time evolution of the center of the Wannier function for each path is plotted in Fig.~\ref{fig:3site_deformpath}(b). 
The number of net charges passing through the unit cell, i.e., $C_{1}^{\{t-k\}}=2$, is indeed the same for all the chosen paths. 
%

% %%%%%%%%%%%%%%%%%%%%%%%
% %%%%%%  Figure 3 %%%%%%
% %%%%%%%%%%%%%%%%%%%%%%%
% %
% \begin{figure}[t]
% \centering
% %
% \includegraphics[width=\linewidth]{./figs/03_trivial_3site_case.png}
% %
% \caption{
% %
% Three-site tight-binding model with a trivial quantum pumping.
% %
% (a) The unit cell of the model with the in-phase onsite potential.
% %
% (b) The time evolution of the Wannier function for the same three parametric paths $\alpha(t)$ as in Fig.~\ref{fig:3site_nontrivial}(c) . 
% }
% \label{fig:3site_trivial_case}
% \end{figure}

To compare with the above nontrivial case, we construct another similar Hamiltonian with a different onsite potential form:
%$H(\alpha)=-2\Delta\sum_{i}\left[\cos(2\pi\alpha)+\frac{3}{2}(x_{i}-\frac{2}{3})\right]c_{i}^{\dagger}c_{i}+\gamma\sum_{i}\left[c_{i+1}^{\dagger}c_{i}+\rm{c.c.}\right].$
%shown in Fig.~\ref{fig:3site_trivial_case}(a), 
%
\begin{equation}\label{eq:trivial_H}
\begin{aligned}
    H(\alpha) =-2&\Delta\sum_{i}\left[\cos(2\pi\alpha)+\frac{3}{2}(x_{i}\%1-\frac{2}{3})\right]c_{i}^{\dagger}c_{i} \\ &+\gamma\sum_{i}\left[c_{i+1}^{\dagger}c_{i}+\rm{c.c.}\right],
\end{aligned}
\end{equation}
%
%to demonstrate that a periodically-driven system with a trivial Chern number $C^{\{\alpha-k\}}$ in the adiabatic limit cannot be classified into different groups.
%
where $\%$ is the modulo operator.
%The only difference between the current case and the previous one is the onsite potential form.
%
In this case, the onsite potential of each site oscillates with the same phase and amplitude; $\Delta=0.2$ and $\gamma=-1$ are chosen to make the system have a trivial Chern number $C^{\{\alpha-k\}}_{n}=0$.
We use the same three parametric paths $\alpha(t)$ as in the previous nontrivial case [see Fig.~\ref{fig:3site_nontrivial}(c)] to compute the time evolution of the center of the Wannier function for the lowest band, which shows no charge pumping $C_{1}^{\{t-k\}}=0$ [see Fig.~\ref{fig:3site_nontrivial}(d)]. 
%
%shown in Fig.~\ref{fig:3site_trivial_case}(b).
%
%All the paths give the same trivial Chern number $C^{\{t-k\}}$, i.e., no charge pumping. 
%
%Therefore, due to the vanishing Chern number $C^{\{t-k\}}$, we cannot classify this time-periodic system.
%
The results from the nontrivial and trivial case imply that the Chern number $C^{\{\alpha-k\}}$ is the building block for the Chern number $C^{\{t-k\}}$.
%, when we use time as a synthetic dimension and the parameter $\alpha$ is a function of time.

%%%%%%%%%%%%%%%%%%%%%%%
%%%%%%  Figure 4 %%%%%%
%%%%%%%%%%%%%%%%%%%%%%%
%
\begin{figure}[b]
\centering
\includegraphics[width=\linewidth]{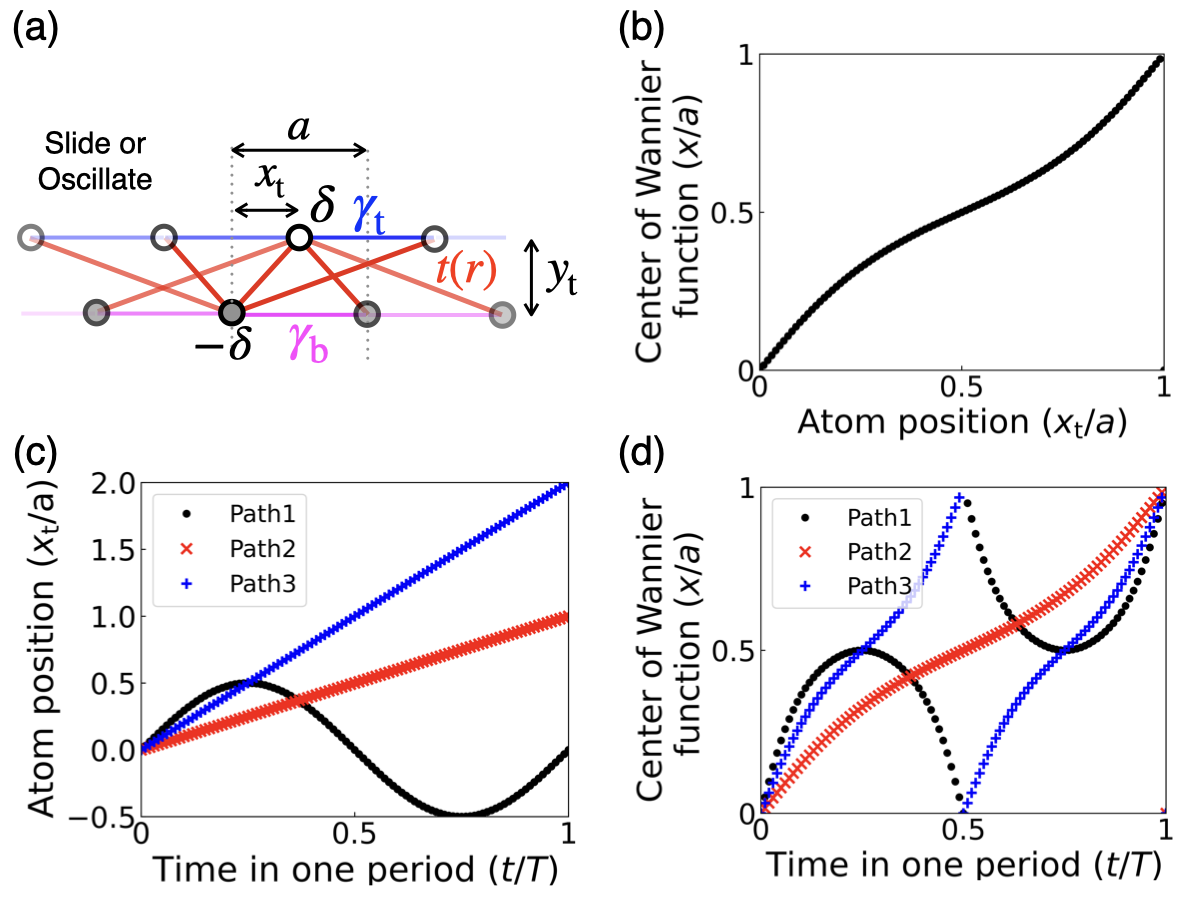}
\caption{
Two-1D-parallel-chains tight-binding model with the upper chain sliding and the lower one fixed.
%
% insulator
(a) The unit cell is shown using the dash lines. The onsite potentials of the two atoms in the unit cell are $-\delta$ and $\delta$, respectively. 
The other parameters are described in the main text. 
The atomic position for the lower chain in the unit cell is $(0, 0)$, while that for the upper one is $(x_{t},y_{t})$. 
(b) The evolution of the Wannier function as the upper chain moves through a unit cell.
(c) The evolution of the atom position $x_{\rm{t}}$ for three paths. Path1: $x_{t}=\sin(2\pi t/T)$; Path2: $x_{t}=t/T$; Path3: $x_{t}=2t/T$.
(d) The evolution of the center of the Wannier function for each path shown in (c). 
}
\label{fig:2chain_model_insulator}
\end{figure}

\subsection*{Two-chains tight-binding model}

The second example consists of two parallel 1D chains with the upper chain sliding with respect to the fixed lower one as shown in Fig.~\ref{fig:2chain_model_insulator}(a). 
This model is inspired by recent experiments where interlayer sliding has been demonstrated in van der Waals crystals and carbon nanotubes ~\cite{viznerstern.waschitz.ea_2021,wu.li_2021,cumings.zettl_2000}.
Compared to the previous example where the ions or sites are fixed, we demonstrate here that the formula [Eq.~(\ref{eq:ctk2cak})] can also be applied to those cases where the ions are moving.
The Hamiltonian with a parametric variable $x_{t}$$-$ the $x$ coordinate of the atom in the upper(top) chain$-$can be written as~\footnote{The two-1D-sliding-chains model includes the possibility of inducing a topological phase transition through parametric sliding. Here we focus on one insulating phase without gap closure during sliding.}
%
% tight binding model for the t%wo-chain model
\begin{equation}\label{eq:2chain_model}
\begin{aligned}
H(x_{t})&=\sum_{j=t,b}\delta_{j}\sum_{i}c_{j,i}^{\dagger}c_{j,i}+\sum_{j=t,b}\gamma_{j}\sum_{i}(c_{j,i}^{\dagger}c_{j,i+1}+\rm{c.c.})
\\ &+ \sum_{ij}\left\{t[r_{(t,i),(b,j)}(x_{t})]c_{t,i}^{\dagger}c_{b,j}+\rm{c.c.}\right\},
\end{aligned}
\end{equation}
where the first line describes the onsite potential and the intra-chain couplings, and the second line the inter-chain couplings. 
Here $\delta_{j}$ is the onsite potential for each site of the bottom ($b$) or top ($t$) chain, and $\gamma_{j}$ the intra-chain coupling for the nearest neighbors. 
%
% Introduction to the Hamiltonian
%
For the inter-chain coupling $t(r)$, which depends on the distance $r$ between one atom in the bottom chain and another one in the top chain, we use
$t(r) = t_{\rm{max}}e^{-(r-y_{\rm{t}})^{2}/2\sigma^{2}}$,
where $t_{\rm{max}}$ is the maximum coupling when the atom in the upper chain is just on top of the one in the lower chain, $y_{t}$ the $y$ coordinate of the atom in the upper chain, and $\sigma$ the effective coupling range.
The distance $r_{(t,i),(b,j)}(x_{t})$ can be evaluated using $\{[x_{t}+(j-i)a]^{2}+y_{t}^{2}\}^{1/2}$.
Below we solve the Hamiltonian in $k$-space, and use a $\sigma$ value to consider only the coupling between the nearest sites in the upper and lower chain. 
%
%Note that our model is similar to the Su-Schrieffer–Heeger (SSH) model~\cite{asboth2016short} with an extra coupling between the next nearest neighbors; therefore, we expect some of the features of the SSH model to show up (see below).

% setup 1): always insulating state
%
The system is set up to remain in the insulating state for any $x_{t}$ value when the upper chain slides or oscillates. 
The parameters for the Hamiltonian are $\gamma_{\rm{b}}=-2.0$, $\gamma_{\rm{t}}=-0.3$, $t_{\rm{max}}=-4.0$, $\sigma=0.5$, and $\delta=0$.
The evolution of the center of the Wannier function for $x_{t}$ is shown in Fig.~\ref{fig:2chain_model_insulator}(b), and the number of net charges pumped within one parametric loop is $1$, i.e., the Chern number for the lowest band.
Notice that in this case, we do not directly compute the Chern number because of the basis change in the tight-binding model due to the movement of the upper atom in the unit cell; instead, we use the number of net pumped charges to represent the Chern number.  

We choose three arbitrary paths for $x_{\rm{t}}$ as shown in Fig.~\ref{fig:2chain_model_insulator}(c). 
Path$1$ corresponds to the top chain oscillating around $x_{\rm{t}}=0$.
Path$2$ and $3$ correspond to the top chain sliding with respect to the bottom chain with a velocity of $1$ and $2$ (in the unit of $a/T$), respectively. 
%
%Note that Path$2$ is the same as the parametric variable $x_{\rm{t}}$ loop, because of $x_{t}=t$.
%
The time evolution of the associated Wannier center for each path is shown in Fig.~\ref{fig:2chain_model_insulator}(d).
Within one  $T$-period, the numbers of net pumped charges in a unit cell for Path$1$, $2$, and $3$ are $0$, $1$, and $2$, respectively.
Again, the Chern number on the $t-k$ plane is a multiple of that on the $x_{t}-k$ plane.

Here we outline a few potential future directions.
Our formula [in Eq.~(\ref{eq:ctk2cak})] can also be extended to include higher spatial dimensions, together with time as another dimension.
In higher spatial dimensions, a periodically-driving system can have relative motions including rotation besides sliding, e.g., rotating bilayer materials~\cite{ma.datta.ea_2022}.
Regarding experimental realizations, the list of parametric variables in a Hamiltonian is not restricted to the above two cases, onsite-potentials and atomic positions; one can also choose an external time-dependent field coupled to the Hamiltonian.

%\section{Conclusion}\label{sec:end_sec}
% future work
In summary, we have developed a general formula to show that the synthetic Chern number involving the time dimension is a multiple of another more fundamental Chern number obtained from a parametric variable $\alpha$.
%
%To use time as a synthetic dimension in evaluating the associated Chern number, we have derived a formula to connect the time-domain Chern number to a multiple of another more fundamental Chern number defined in a space including the parametric variable that directly couples to time. 
%
The integer (multiple) that connects the two Chern numbers can be used to classify a periodically-driven system in the adiabatic limit.
%
%We have validated the formula using two tight-binding models. 
%
% Lastly, here we provide a recipe  is 
% %
% 1) to find a parametric variable for the Hamiltonian, 
% 2) to find the `Brillouin Zone' for the parameter, 
% 3) to couple time to the parametric variable, and 4) to promote time to a synthetic dimension. 
% %
% The Chern number on the $t-k$ plane will be a multiple of the Chern number on the $\alpha-k$ plane.
% %
% The multiple (integer) can thus be used to classify the motion of the periodically-driven system. 
% %
To sum up, our formula provides a tool to investigate and classify the topological properties for periodically-driven systems that can be applied within the adiabatic approximation. 
% %
% The result can be easily generalized to multi-parametric time-dependent periodically-driven Hamiltonian.
% %
% We expect that in this case the Chern number obtained using time as a synthetic dimension would be some products of the Chern numbers that are evaluated in the parametric variables, which would be the subject of future work.
% %

%%%%%%%%%%%%%%%%%%%%%%%%%%%%%%%%%%%%%%%%%%%%%%%%%
% Acknowledgements
%%%%%%%%%%%%%%%%%%%%%%%%%%%%%%%%%%%%%%%%%%%%%%%%%
%
\begin{acknowledgments}
\vspace{10pt}
We acknowledge financial support from the European Research Council (ERC-2015-AdG-694097). 
The Flatiron Institute is a division of the Simons Foundation.
This work was supported by the Cluster of Excellence Advanced Imaging of Matter (AIM), Grupos Consolidados (IT1453-22) and SFB925.
I-T. Lu and D. Shin thank Alexander von Humboldt Foundation for the support from Humboldt Research Fellowship. 
J. Zhang acknowledges funding received from the European Union’s Horizon 2020 research and innovation program under the Marie Sklodowska-Curie grant agreement No. 886291 (PeSD-NeSL).
The authors thank Dr. Marios Michael, Dr. Ofer Neufeld, Dr. Shunsuke Sato, Dr. Peizhe Tang, Anatoly Obzhirov, and Osamah Sufyan for the fruitful discussion.
%
%This research used resources of the National Energy Research Scientific Computing Center, a DOE Office of Science User Facility supported by the Office of Science of the U.S. Department of Energy under Contract No. DE-AC02-05CH11231.
%
\end{acknowledgments}

\bibliography{reference_short_name}

\end{document}